\documentclass[12pt,twoside]{article}
\usepackage{xrb2007}
\usepackage{psfig}
\begin{document}

\session{Jets}
\session{Extragalactic Populations}

\shortauthor{Kundu, Maccarone, \& Zepf}
\shorttitle{LMXBs and Globular Clusters}

\title{The Low Mass X-ray Binary - Globular Cluster Link and its Implications}
\author{Arunav Kundu, Stephen E. Zepf}
\affil{Dept. of Physics \& Astronomy, Michigan State University, East Lansing, MI 48824-2320 }
\author{Thomas J. Maccarone}
\affil{School of Physics \& Astronomy, University of Southampton, Southampton, UK SO17 1BJ }

\begin{abstract}
Studies of nearby elliptical and S0 galaxies reveal that roughly half of the low mass X-ray binaries (LMXBs), which are luminous tracers of accreting neutron star or black hole systems, are in clusters. There is a surprising tendency of LMXBs to be preferentially associated with metal-rich globular clusters (GCs), with metal-rich GCs hosting three times as many LMXBs as metal-poor ones. There is no convincing evidence of a correlation with GC age so far. In some  galaxies the LMXB formation rate varies with GC color even within the metal-rich peak of the typical bimodal cluster metallicity distribution. This provides some of the strongest evidence to date that there are metallicity variations within the metal-rich GC peak, as is expected in hierarchical galaxy formation scenarios. We also note
that apparent correlations between the interaction rates in GCs and LMXB frequency may not be reliable because of the uncertainties in some GC parameters. We argue in fact that there are considerable uncertainties in the integrated properties of even the Milky Way clusters that are often overlooked.
\end{abstract}

\section{Introduction}
Chandra X-ray images of nearby ellipticals and S0s reveal large numbers of point sources, confirming a long-standing suggestion that the hard X-ray emission in many of these galaxies comes predominantly from X-ray binaries. In the old stellar populations of these galaxies the bright, L$_X$$\geq$10$^{37}$ erg s$^{-1}$ sources detected in typical Chandra observations are low mass X-ray binaries, binary systems comprising a neutron star or black hole accreting via Roche lobe overflow from a low mass companion.  

An interesting characteristic of LMXBs is that they are disproportionately abundant in globular clusters. In the Milky Way (MW) the probability of finding a LMXB among the stars in the globular clusters is a few hundred times larger than for field stars. This has long been attributed to efficient formation of LMXBs in GCs due to dynamical interactions in the core. Early type galaxies are ideal for studies of the LMXB-GC link as they are particularly abundant in clusters. While there are $\approx$150 globular clusters in the Milky Way, bright elliptical galaxies typically host several thousand GCs. Globular clusters are among the simplest stellar systems known; Thus the stellar population in each of these compact, $\sim$10$^5$ M$_\odot$, luminous systems has well defined characteristic properties such as age and metallicity.  The identification of LMXBs with GCs provides a unique opportunity to probe the effects of these parameters on LMXB formation and evolution. 

\begin{figure}[!ht]
\begin{center}
\centerline{\psfig{figure=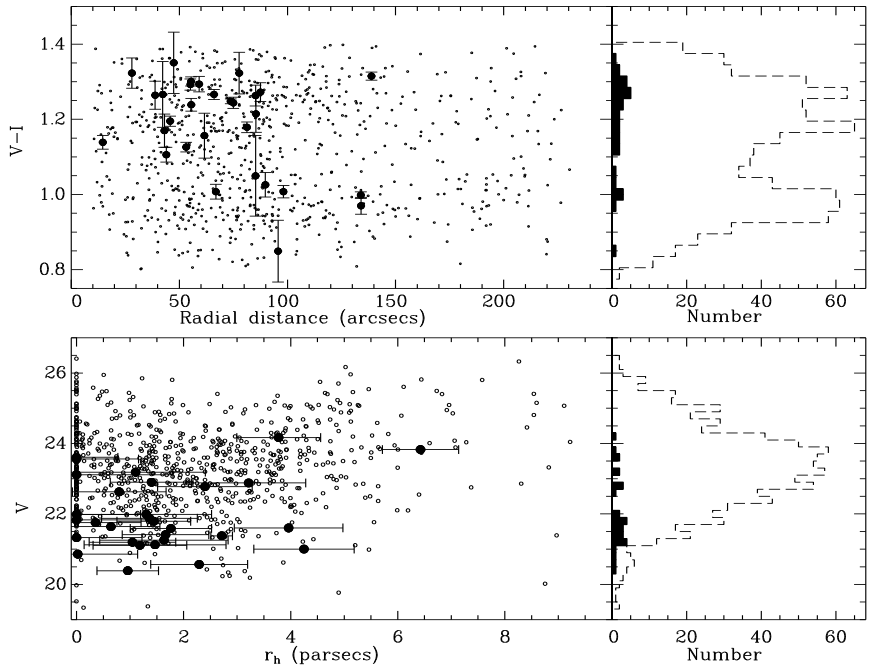,width=4.5in}}
\caption{
Top: The V-I colors of GCs vs. distance from the center of NGC 4472 and GC color distribution. LMXB-GC matches are indicated by large filled circles/bins. The colors of these clusters primarily trace the underlying (bimodal) metallicity distribution. Bottom: V magnitude of globular clusters vs. half light radius and the globular cluster luminosity function. LMXBs are preferentially located in the high mass and metal-rich GCs. There is a weak anti-correlation with GC half-light radius, but no obvious correlation with galactocentric distance. Each of these trends has been confirmed in other galaxies. 
}
\end{center}
\end{figure}

\section{Globular Cluster - LMXB correlations and the implications on LMXB, Globular Cluster, and Host Galaxy Formation \& Evolution  }

The unique spatial resolution of the HST in the optical is an ideal compliment to Chandra for LMXB-GC studies because it resolves out individual  clusters even in the innermost regions of galaxies. Figure 1 plots the colors, magnitudes, half-light radii, and galactocentric distances of globular clusters identified in a HST based study of NGC 4472 \citep[][KMZ02]{2002ApJ...574L...5K}, the most luminous elliptical galaxy in the Virgo cluster. The well known Gaussian globular cluster luminosity function and  bimodal GC color distribution (which primarily traces the underlying  metallicity distribution) are apparent. The large symbols indicate GCs that host LMXBs. Clearly LMXBs are preferentially found in the higher mass and metal-rich GCs. On the other hand statistical tests (KMZ02) reveal only a marginal tendency for LMXBs to favor GCs with smaller half light radius, and no convincing correlation with galactocentric distance. These trends with GC properties have subsequently been confirmed in other galaxies \citep[][KMZ07]{2006ApJ...647..276K,2007ApJ...660.1246S,2007ApJ...662..525K}. Overall $\approx$45\% of the bright LMXBs observed in typical Chandra observations of early type galaxies are located in GCs. However, the recent studies of KMZ07 and \citet{2007arXiv0710.5126F} suggest that the fraction may be smaller at lower X-ray luminosities  raising intriguing questions about differences in the nature of the LMXBs in the field and in clusters.

\begin{figure}[!h]
\begin{center}
\centerline{\psfig{figure=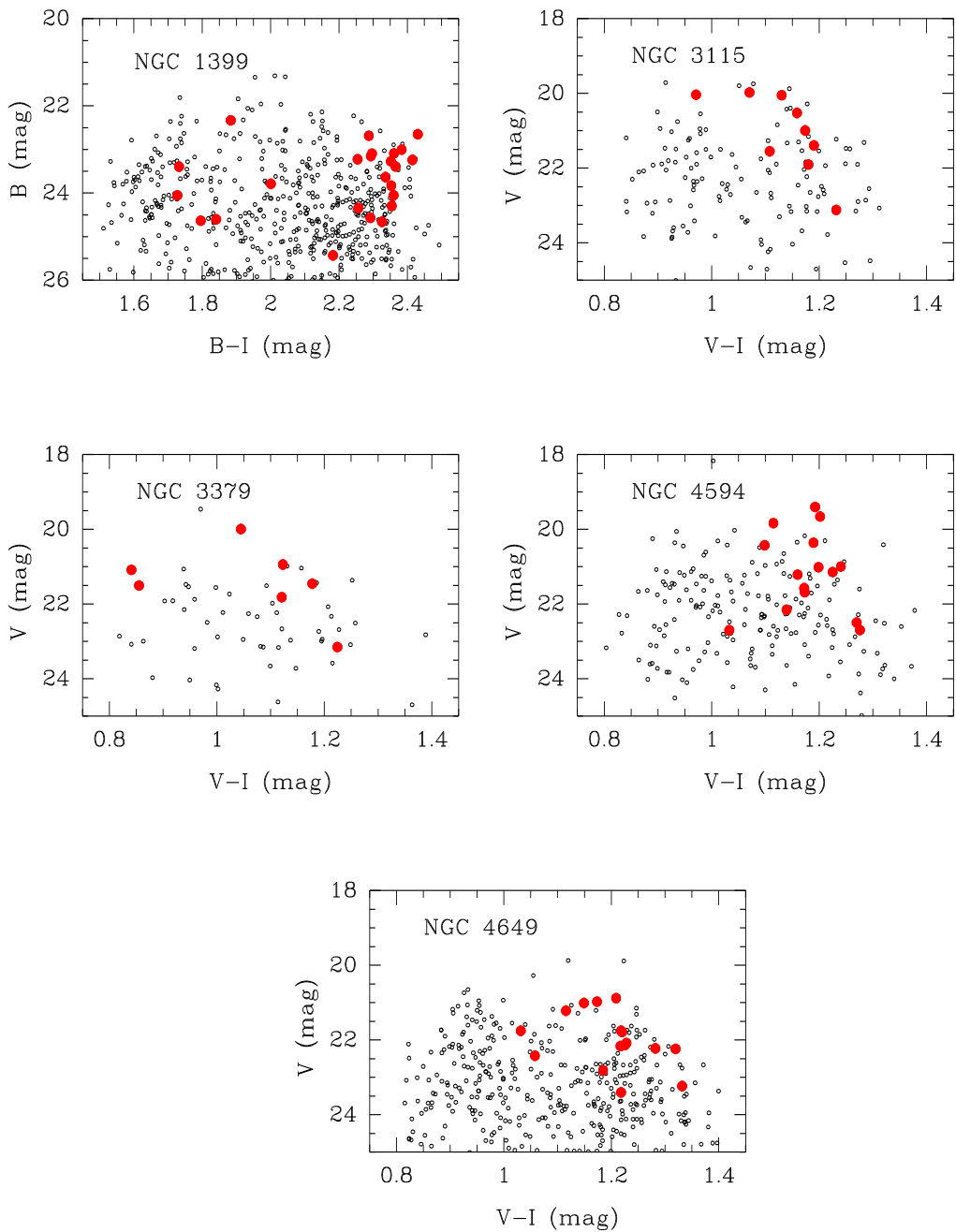,width=5.3in}} 
 \caption{Color-magnitude diagrams for globular clusters in five elliptical galaxies with known bimodal cluster metallicity (color) distributions. Large filled points represent GCs with LMXB counterparts. LMXBs are clearly found preferentially in luminous (high mass) and red (metal-rich) globular clusters. There is an obvious enhancement of LMXBs in the reddest GCs in the metal-rich cluster subpopulation of NGC 1399. This provides some of the strongest evidence to date that there is metallicity substructure within the red GCs in this giant elliptical galaxy, as expected in hierarchical merging scenarios of galaxy formation.}
\end{center}
\end{figure}

	These correlations provide a window the physics of LMXBs, and the dynamics of globular clusters. High mass clusters in the Local Group galaxies are known to be denser than their low mass counterparts.  Since luminous clusters also obviously contain more stars the higher rate of LMXB formation in high mass GCs can be explained by the higher interaction rate in such clusters. However we note that the mass of a cluster is highly correlated to the interaction rate, and we argue below that the apparent core interaction rate calculated in some studies likely does not contain any additional information due to the uncertainties in various measured GC parameters.

There is no obvious dynamical reason for the three times larger rate of LMXBs in the red, metal-rich, GCs in elliptical galaxies as compared to the metal-poor ones. It is known that there is a similar excess of bright LMXBs in the metal-rich GCs of the Milky Way. However in the past this was often dismissed as the consequence of the location of these clusters in the Galactic bulge, where the dynamical evolution of GCs may be accelerated. The lack of correlation of LMXB frequency in GCs with galactocentric distance in the large GC samples of ellipticals (KMZ02; KMZ07) argues against this interpretation and establishes a primary correlation with GC color.

	It has also been suggested that the enhanced rate of LMXBs in metal-rich GCs may be because of the younger ages of these clusters. However infrared imaging, which yields more accurate constraints on the metallicities of GCs and helps  disentangle the age-metallicity degeneracy, has established that the optical colors of globular cluster systems that have clearly bimodal distributions are primarily a tracer of metallicity \citep{2007ApJ...660L.109K} and LMXBs are indeed preferentially located in metal-rich GCs \citep{2003ApJ...589L..81K,2007ApJ...661..768H}. Fig 2 shows the color magnitude distributions of five galaxies with confirmed bimodal cluster color distributions and the factor of three enhancement with metallicity.

The color distribution of GCs hosting LMXBs in NGC 1399 reveals a very interesting feature: Not only is there a larger fraction of LMXBs in the red, metal-rich sub-population, but in fact LMXBs are preferentially located in the  reddest GCs. This is the most convincing evidence to date that there is metallicity structure within the metal-rich peak of GCs, as is expected from hierarchical models of galaxy and globular cluster system formation. Such structure in the metallicity distribution is obviously difficult to identify in integrated light studies of galaxies, and even optical studies of the simple stellar systems of GCs have not clearly identified such a feature. At present it is not clear if the reason that this effect is obvious only in NGC 1399 is because of the location of this galaxy at the center of the Fornax cluster which leads to a particularly efficient enrichment history, or because of the larger color baseline of this data set. Although the Sivakoff et al. (2007) analysis is broadly in agreement about the metallicity trend and finds a linear increase in the fraction of LMXBs in GCs with metallicity, it suggests that the rate of LMXBs in the reddest, and consequently most metal-rich GCs, actually drops. We note that LMXBs are found preferentially in the brightest clusters, and if the GC sample in Sivakoff et al. (2007) were to be restricted to similar luminosities this apparent discrepancy would disappear. In other words the reddest GCs in the Sivakoff et al. (2007) sample are the faint ones which have scattered to these colors due to observational uncertainties and are not representative of the true underlying color/metallicity of the cluster distribution.

	There have been some theoretical attempts to explain the metallicity effect.  In \citet{2004ApJ...606..430M} we suggest that irradiation of the donor star by the LMXB is key. Higher metallicity objects can dissipate this energy through line cooling while a wind is generated in low metallicity star, thus lowering the LMXB lifetime. As a consequence of the wind the X-ray emission from low metallicity LMXBs should be harder because of the excess emission, a feature that has been observed in NGC 4472 \citep{MKZ03} and M31 \citep{IB99}.  \citet{2006ApJ...636..979I} on the other hand suggests that the absence of an outer convective layer in solar mass metal-poor stars limits magnetic braking and the formation of mass transferring LMXBs.

\section{Uncertainties in the Integrated Properties of Globular Clusters in the Milky Way and Other Galaxies and the Impact on Interaction Rates }

\begin{figure}[!h]
\begin{center}
\centerline{\psfig{figure=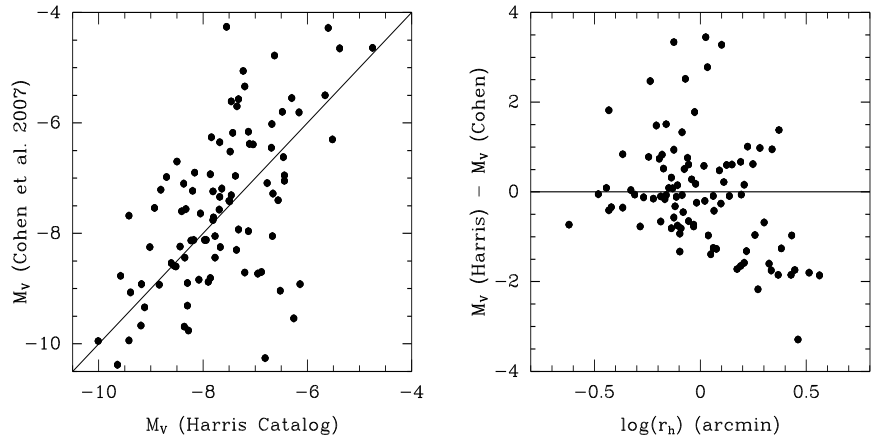,width=4.5in}} 
 \caption{Left: The absolute magnitude of MW GCs from the Harris (1996) compilation compared to the measurements in Cohen et al. (2007). There is considerable scatter at all magnitudes. Right: The difference between the magnitudes in the Harris compilation and Cohen et al. (2007) values as a function of the measured half-light radius in arcmins {\it on the sky}. Cohen et al. (2007) measured the light within 50$''$ for all clusters. If the difference in the two sets is due to the limited spatial coverage of the Cohen et al. (2007) data there should be a strong correlation in this figure. The lack of a correlation suggests that there is considerable uncertainty in the integrated magnitudes of MW globular clusters.   }
\end{center}
\end{figure}

\begin{figure}[!h]
\centerline{\psfig{figure=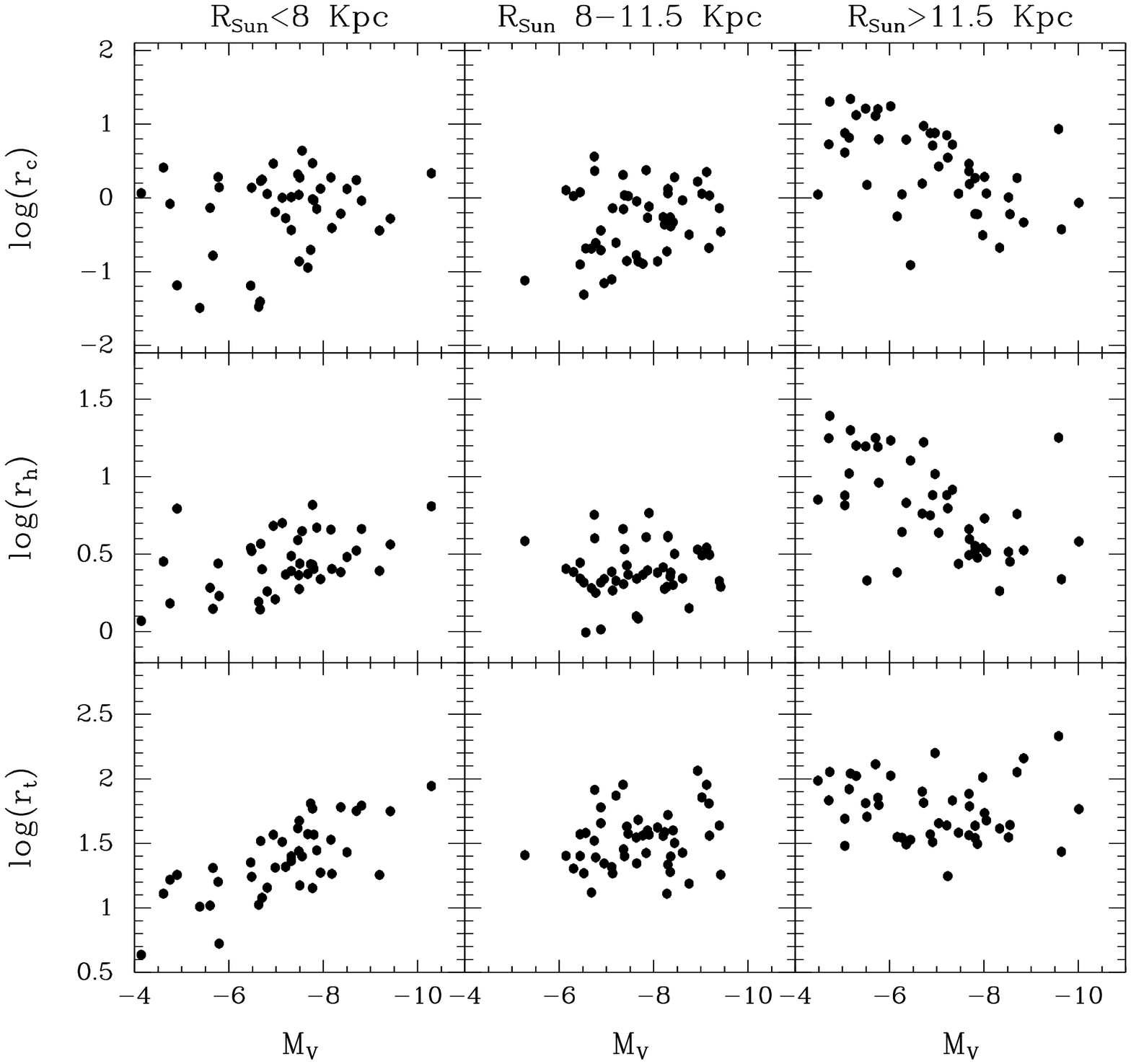,width=3.9in}} 
\vspace*{0.2 cm}
\centerline{\psfig{figure=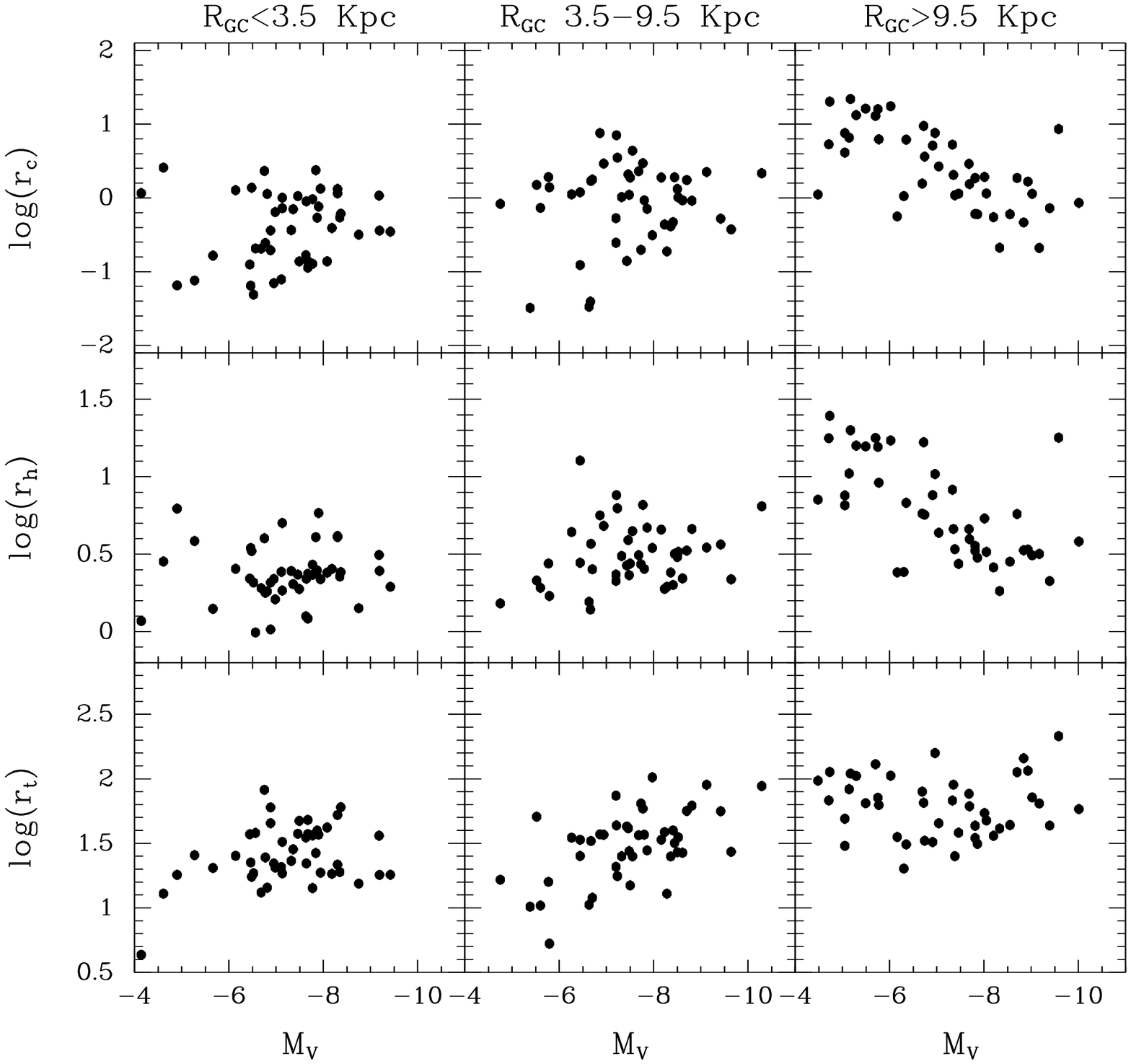,width=3.9in}} 
 \caption{Top: The variation of the core, half-light, and tidal radii of MW globular clusters with absolute magnitude for three ranges of Heliocentric distance. Bottom: The corresponding plots for three galactocentric distance ranges. The direction of the correlation for all three size measurements appears to change with Heliocentric distance. As this effect is weaker in the lower plot it argues either that the properties of MW clusters are somehwat unusual in the Solar neighborhood, or more likely there are observational biases in the commonly accepted values of the MW globular cluster parameters. }
\end{figure}

There have been attempts to derive the dependence of the observed LMXB rate on the globular cluster metallicity and mass by assuming a specific M/L ratio and color-metallicity correlation by us and other groups \citep{2004ApJ...613..279J,2006A&A...458..477S,2007ApJ...660.1246S}. These studies generally agree that the mass dependence is roughly linear and the metallicity dependence goes as approximately Z$^{0.25}$. Jordan et al. (2004) and Sivakoff (2007) further attempt to link the probability of finding a LMXB in a GC to the rate of stellar interactions in the core of a GC (Also see the contribution by Sivakoff elsewhere in this volume). However, this requires measuring the core radii of GCs which is a fraction of even a HST pixel for these galaxies. Thus the core radius is derived by extrapolation of other measured GC properties. It is not clear if any of these optical data sets have sufficient depth to simultaneously measure three parameters for each GC, the total magnitude, at least one fiducial radius (core, half-light, or tidal) and the shape of the light profile. Smits et al. (2006) show that the Jordan et al. (2004) results are as expected if there is no information about the core radius of the clusters in their sample. In \citet{K08} we show that there are biases in the measured properties of even such basic properties of GCs like the total magnitude in many published HST studies of globular cluster systems that do not fully account for the resolving power of the HST. In fact we argue below that even the integrated properties of Milky Way GCs likely have much larger uncertainties than assumed, which has significant impact to the derived interaction rates, and hence the LMXB formation efficiency.

Recent studies have attempted to isolate the effects of the interaction rate and cluster mass of Milky Way GCs on the rate of close binaries such as LMXBs \citep{P03,PH06}. Other contributions in this volume study the effects of interaction rates, or other Milky Way GC parameters, in subsamples  of clusters. These studies typically adopt the integrated properties of the Milky Way clusters from the \citet{H96} compilation. But the Milky Way clusters have been analyzed in innumerable studies using different instruments and measurement standards over the years. Many of the commonly accepted values of cluster parameters come from many generations of compilations of, in cases, decades old data. Just how reliable are these numbers?

In a recent study \citet{C07} analyzed the optical and IR properties of Milky Way GCs. In the course of this analysis they re-measured the V-band magnitude within a 50$''$ aperture for the GCs in their sample by integrating the surface brightness profile. The left panel of Figure 3, which compares the absolute magnitudes of Milky Way GCs from the Cohen et al. (2007) study to the Harris (1996) compilation reveals considerable scatter at all magnitudes. The lack of a strong correlation between the difference in magnitudes of the two compilations with the size of a GC on the sky indicates that the {\it considerable} scatter cannot be explained by the limited spatial coverage of the Cohen et al. (2007) analysis.

There are hints about problems elsewhere. In Fig 4 we plot the variation of the core, half-light, and tidal radii of MW GCs with M$_V$  for three ranges of Heliocentric and Galactocentric distances (roughly equal number of GCs in each set).  The direction of the correlation for all three size measurements appears to change with Heliocentric distance. The lack of a strong trend with Galactocentric distance argues either that the properties of MW clusters are somewhat unusual in the Solar neighborhood, or more likely there are observational biases (possibly associated with the size of the object in the sky among other things) in the commonly accepted values of the MW GC parameters. This has significant impact not only on GC dynamics and LMXB formation rates but more generally on globular cluster research.

\end{document}